%% file: main.tex
\documentclass[conference]{IEEEtran}

\usepackage{amsmath,amssymb,amsfonts}
\usepackage{algorithmic}
\usepackage{array}
\usepackage{cite}
\usepackage{hyperref}
\usepackage{graphicx}
\usepackage{subfiles}
\usepackage{tabularx}
\usepackage{textcomp}
\usepackage{xcolor}
\usepackage{subfig}
\usepackage{xcolor}

\hypersetup{
  hidelinks,
  pdftitle={Hybrid Deep Neural Network for Brachial Plexus Nerve Segmentation in Ultrasound Images},
  pdfauthor={},
  pdfsubject={},
  pdfkeywords={{Medical Imaging}, {Brachial Plexus},  {Deep Learning}, {Convolutional Neural Network},{Segmentation}}
}

\def\BibTeX{{\rm B\kern-.05em{\sc i\kern-.025em b}\kern-.08em
    T\kern-.1667em\lower.7ex\hbox{E}\kern-.125emX}}
\begin{document}

\title{Hybrid Deep Neural Network for Brachial Plexus Nerve Segmentation in Ultrasound Images}

\author{\IEEEauthorblockN{Juul van Boxtel \IEEEauthorrefmark{3}}
\IEEEauthorblockA{\textit{Department of Biomedical Engineering} \\
\textit{Eindhoven University of Technology}\\
Eindhoven, the Netherlands \\
\href{mailto:j.p.a.v.boxtel@student.tue.nl}{j.p.a.v.boxtel@student.tue.nl}}
\\
\IEEEauthorblockN{Josien Pluim}
\IEEEauthorblockA{\textit{Department of Biomedical Engineering} \\
\textit{Eindhoven University of Technology}\\
Eindhoven, the Netherlands \\
\href{mailto:J.Pluim@tue.nl}{J.Pluim@tue.nl}}
\and
\IEEEauthorblockN{Vincent Vousten \IEEEauthorrefmark{3}}
\IEEEauthorblockA{\textit{Department of Biomedical Engineering} \\
\textit{Eindhoven University of Technology}\\
Eindhoven, the Netherlands \\
\href{mailto:v.r.j.vousten@student.tue.nl}{v.r.j.vousten@student.tue.nl}}
\\
\IEEEauthorblockN{Nastaran Mohammadian Rad}
\IEEEauthorblockA{\textit{Department of Biomedical Engineering} \\
\textit{Eindhoven University of Technology}\\
Eindhoven, the Netherlands \\
\href{mailto:n.mohammadian.rad@tue.nl}{n.mohammadian.rad@tue.nl}}}

\maketitle
\begingroup\renewcommand\thefootnote{\IEEEauthorrefmark{3}}
\footnotetext{These authors contributed equally to this work.}

\begin{abstract}
\subfile{chapters/abstract}
\end{abstract}


\section{Introduction}\label{section:introduction}
\subfile{chapters/introduction}

\section{Methods}\label{section:methods}
\subfile{chapters/methods}

\section{Experiments}\label{section:experiments}
\subfile{chapters/experiments}

\section{Results}\label{section:results}
\subfile{chapters/results}

\section{Discussion}\label{section:discussion}
\subfile{chapters/discussion}

\section{Conclusion}\label{section:conclusion}
\subfile{chapters/conclusion}

\bibliographystyle{IEEEtran}
\bibliography{references}\phantomsection

\end{document}

%% file: chapters/abstract.tex
Ultrasound-guided regional anesthesia (UGRA) can replace general anesthesia (GA), improving pain control and recovery time. This method can be applied on the brachial plexus (BP) after clavicular surgeries. However, identification of the BP from ultrasound (US) images is difficult, even for trained professionals. To address this problem, convolutional neural networks (CNNs) and more advanced deep neural networks (DNNs) can be used for identification and segmentation of the BP nerve region. In this paper, we propose a hybrid model consisting of a classification model followed by a segmentation model to segment BP nerve regions in ultrasound images. A CNN model is employed as a classifier to precisely select the images with the BP region. Then, a U-net or M-net model is used for the segmentation. Our experimental results indicate that the proposed hybrid model significantly improves the segmentation performance over a single segmentation model.

%% file: chapters/introduction.tex
\begin{figure*}[!tb] 
    \centering
    \subfloat[Classification model (CNN) \label{subfig:ClassModel}]{\includegraphics[width=0.65\linewidth]{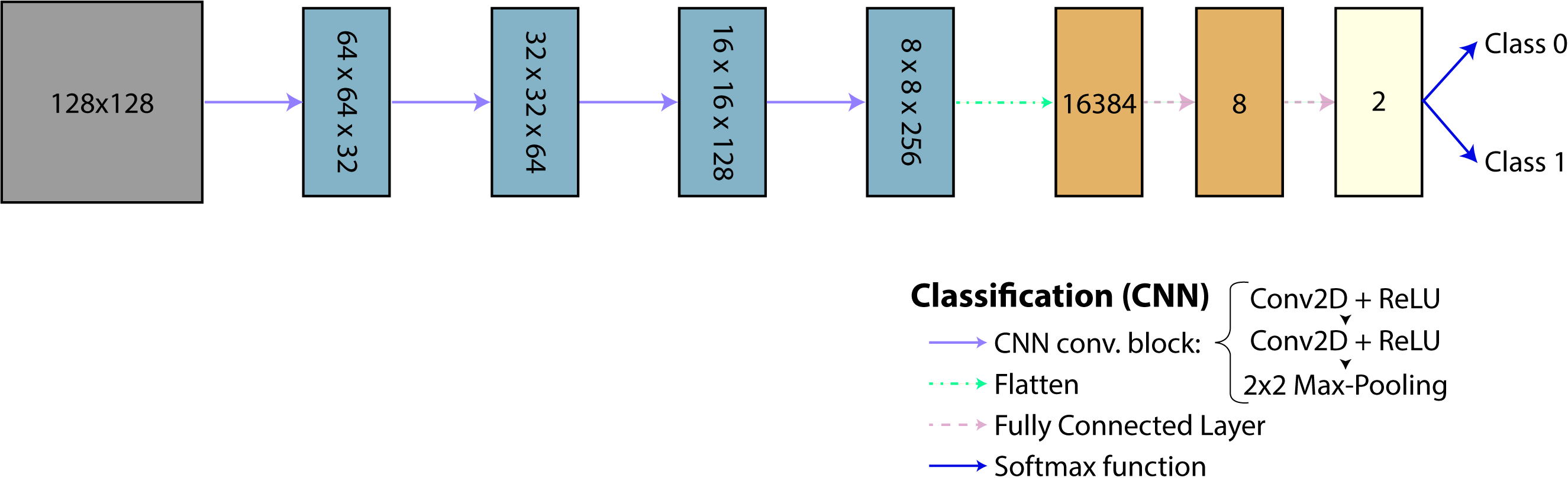}} \\
    \subfloat[Segmentation model (U-net/M-net)  \label{subfig:SegModel}]{\includegraphics[width=0.85\linewidth]{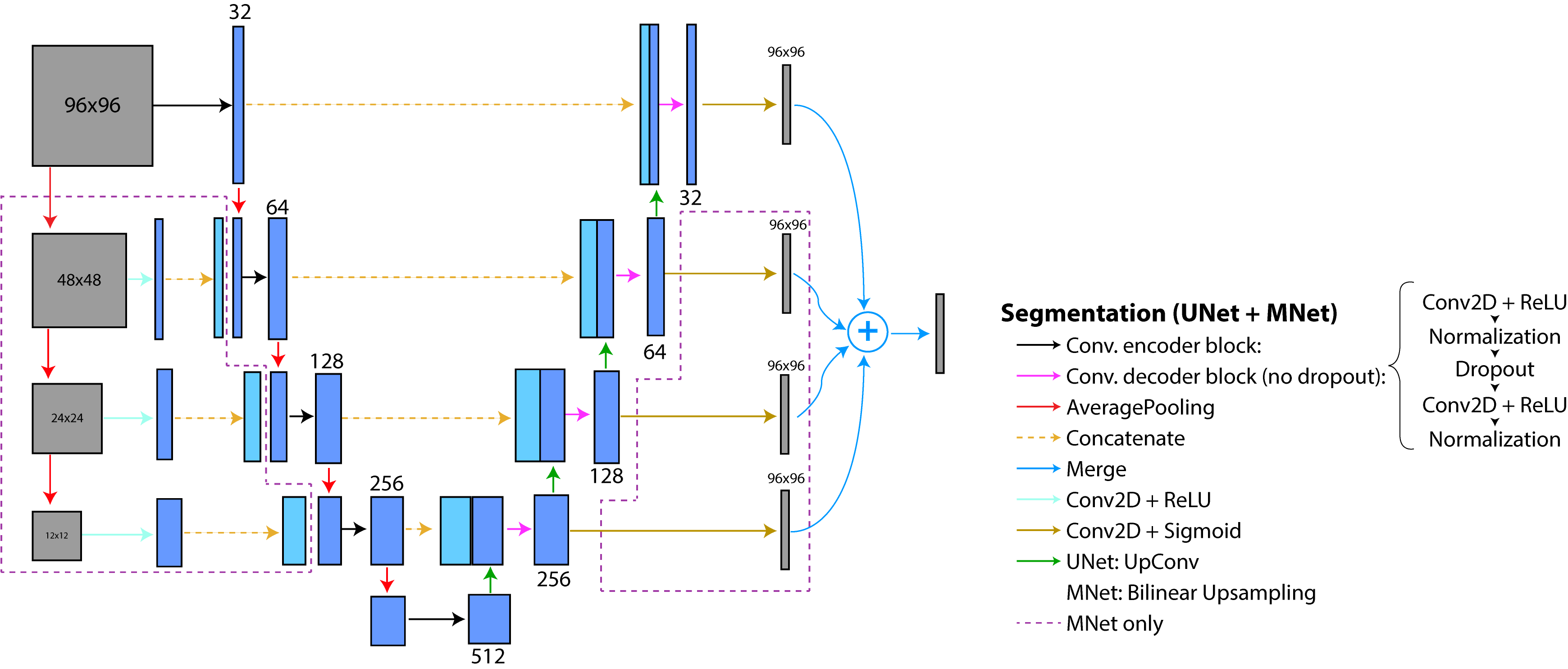}}
    \caption{The architecture of (a) CNN for the classification and (b) U-net and M-net for the segmentation.}
    \label{fig:Hybrid_Model} 
\end{figure*}

The brachial plexus (BP) is a part of the cervical nerves which originates from the spinal cord. It is partially located in the neck and partially in the axilla. It is an important part of the nervous system, as it innervates the upper limb. The BP contains cervical nerves C\textsubscript{5}-C\textsubscript{8} and most of the thoracic T\textsubscript{1} nerves, often along with fibers from C\textsubscript{4} and/or T\textsubscript{2} \cite{Marieb}.
The segmentation of BP regions is an important consideration in treatment planning for lung or head-and-neck cancer patients, as radiotherapy used to treat cancer patients can induce brachial plexopathy when the BP is overburdened with radiation energy, causing severe and irreversible effects \cite{Yang2013}.
Furthermore, the BP is used more for regional anesthesia (RA) in clavicular surgeries, in the form of interscalene BP blocks. It can replace general anesthesia, allowing better pain control, decreasing opioid consumption, and reducing recovery time \cite{Pincus2019}, \cite{Banerjee2019}. The classic RA procedure is to inject the anesthesia into the target nerve region rather blindly. This method poses risks as block failure, nerve trauma, and local anesthesia toxicity \cite{Kakade2018}. Ultrasound-guided RA (UGRA) has become a popular method to visualize this procedure. UGRA is often faster, requires fewer attempts, and in some cases provides a better sensory block compared to other RA techniques \cite{Liu2009}.  

In general, tissue segmentation in ultrasound (US) images is a challenging task due to low contrast between background and the tissue, compared to other modalities like MRI and CT scans.
Especially the segmentation of nerve tissue can be challenging due to speckled noise coming with the US modality and the fact that the nerve region does not form a salient structure in the images \cite{Abraham2019}. The need for trained experts, capable of recognizing the region of interest (ROI), limits the applicability of the UGRA procedure. Deep learning algorithms could provide a solution by automating the recognition of the ROI. This would make UGRA better available for more extensive use.

Deep learning methods have shown promising results in image segmentation applications\cite{Wang2019}. Deep learning algorithms, and more specifically convolutional neural networks (CNNs) and U-net are widely used in medical applications such as tumor region identification and metastasis detection \cite{Seetha2018, Kong2017}.
CNNs are suitable to be applied to data that have a grid-like topology such as time-series and images \cite{Goodfellow-et-al-2016}. A typical CNN has a hierarchical architecture that alternates a convolution layer, a rectified linear unit (ReLU) activation function, and a pooling layer to summarize the large input spaces into a lower-dimensional feature space. CNN solutions are among the best-performing systems on pattern recognition systems\cite{Goodfellow-et-al-2016}.

U-net \cite{UNetArticle} is the leading model architecture for medical image segmentation. The model consists of an encoder and a decoder part. The encoder part, the left side of the U-shape, resembles a normal CNN. It consists of subsequent convolution, activation, and pooling layers. The decoder part, the right side of the U-shape, is symmetrical to the encoder part. It consists of upsampling layers, a concatenating layer which adds the feature map of the encoder layer, and subsequent convolutional layers. The U-net architecture has proven to be able to accurately localize ROIs, even when only trained on relatively small training sets \cite{UNetArticle}. Over the years, a variety of adaptations have been made to the U-net structure in attempts to improve its performance on specific datasets. A promising architecture is M-net \cite{MNetArticle}. It uses multiple scaled inputs and multiple outputs coming from the different layers, in order to densely supervise extracted feature maps. 

The goal of this paper is to investigate the effect of implementing a hybrid model in segmenting the BP nerve region from US images. We hypothesize that a combination of CNN and a segmentation model provides more accurate BP segmentation compared to using only a segmentation model.

The rest of this paper is structured as follows: Section \ref{section:methods} presents our hybrid deep neural network architecture. The experimental materials and the procedures are described in Section \ref{section:experiments}. The results of experiments are described and discussed in Sections \ref{section:results} and \ref{section:discussion}. Section \ref{section:conclusion} concludes this paper by summarizing our achievements.

%% file: chapters/methods.tex
In this paper, we used a CNN architecture comprised of four convolutional layers. The architecture is shown in Figure \ref{subfig:ClassModel}. Each convolutional layer consists of two sliding windows with a 3x3 kernel size and ReLU activation, followed by a 2x2 max-pooling layer. In the first layer, 32 filters were used, and this number was multiplied by a factor two every layer down. The output was then flattened and followed by two fully-connected layers. After the first fully-connected layer, a dropout layer with a dropout fraction of 50\% was used.

The first segmentation model used was an adaptation of the U-net architecture \cite{UNetArticle}. The encoder part of the model consists of five convolutional layers, each consisting of two 5x5 sliding windows with ReLU activation, each followed by a GroupNormalization layer with group size 8. In the first layer, the number of filters is 32. This was multiplied by a factor two every layer down. A dropout layer with a dropout fraction of 40\% was used inside the convolutional block, followed by an average-pooling layer. The decoder uses the same convolutional blocks, however without the dropout layer. Between the blocks, up-sampling is done by using transposed convolutions. The final output is given through a 1x1 convolution with Sigmoid activation.

The other model used for segmentation is the M-net architecture \cite{Abraham2019}. M-net is an extension of the U-net architecture, as displayed in Figure \ref{subfig:SegModel}. The M-net structure uses multiple inputs, which are down-sampled versions of the original input by factor two, four, and eight. These inputs are sent in the subsequent layers from the encoder part of the U-net structure, by concatenating them with the feature map of the previous layer. In the decoder part, each layer creates an output image by sending the feature map through a 1x1 convolution with Sigmoid activation, and then up-sampling that output to match the shape of the original input. All four output images are then averaged to form a single final segmentation map. The inner (U) structure is similar to U-net, starting with 32 filters in the first layer. However, a dropout fraction of 20\% is used and the normalization layers are left out. In the decoder, up-sampling is done by bilinear up-sampling. The hyperparameters used are given in Table \ref{table:Hyperparameters}.

Keras library \cite{Keras} was used in our implementation. The model was trained on a dedicated GPU server, containing an Intel Core i7-5930K CPU, 2 NVIDIA GeForce GTX Titan X (GM200) and 1 NVIDIA GeForce GTX Titan Xp (GP102) GPU and 62 GB of RAM.

\renewcommand{\arraystretch}{1.15}
\begin{table}[tb]
\caption{The value of parameters for CNN, U-net, and M-net architectures.}
\label{table:Hyperparameters}
\centering
\begin{tabular}{m{2.2cm}  m{1.5cm}  m{1.5cm}  m{1.5cm}}
\hline
\textit{Parameter} & \textit{CNN} & \textit{U-net} & \textit{M-net}\\
\hline
\textbf{Optimizer} & Adam & Adam & Adam \\
\textbf{Batch Size} & 64 & 16 & 16 \\
\textbf{Learning rate} & 10\textsuperscript{-4} \textsuperscript{(1)} & 10\textsuperscript{-4} \textsuperscript{(2)} & 10\textsuperscript{-4} \textsuperscript{(2)} \\
\textbf{No. epochs} & 25 & 50 & 50 \\
\textbf{Start no. filters} & 32 & 32 & 32 \\
\textbf{Kernel size} & 3x3 & 5x5 & 3x3 \\
\textbf{Dropout} & 0.5 & 0.4 & 0.2 \\
\hline
\multicolumn{4}{p{7.8cm}}{\textsuperscript{1} The CNN optimization algorithm used the `Reduce Learning Rate on Plateau' callback in the Keras API; when the validation accuracy would not improve for 3 consecutive epochs, the learning rate would be lowered by a factor 0.2. The value in the table represents the starting learning rate.}\\
\multicolumn{4}{p{7.8cm}}{\textsuperscript{2} These models used learning rate decay. The LR of U-net was divided by ten at epoch ten, and then after each fifth epoch. The same goes for M-net, but then for every tenth epoch. The value in the table represents the starting learning rate.} \\

\end{tabular}
\end{table}

%% file: chapters/experiments.tex
\renewcommand{\arraystretch}{1.20}
\begin{table}[tb]
\caption{Number of images in BP and no-BP classes before data augmentation.}
\label{table:Nr_images}
\centering
\begin{tabular}{c|c c}
\hline
\textit{} & \textit{Non-filtered} & \textit{Filtered}\\ \hline
BP & 2322 & 1454 \\ 
No-BP & 3313 & 2452 \\ \hline
\end{tabular}
\end{table}

\begin{figure}[tb] 
    \centering
    \subfloat[Ultrasound Image of the neck\label{subfig:ExampleImage1}]{\includegraphics[width=0.45\linewidth]{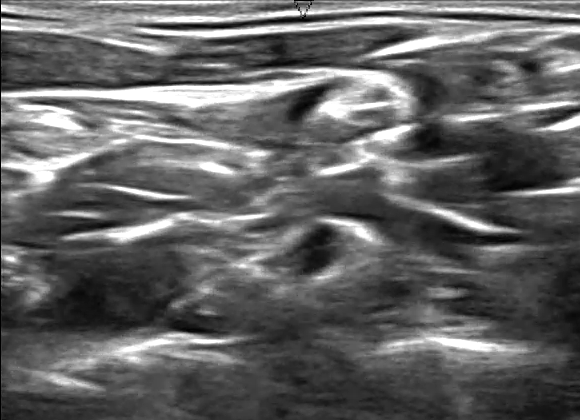}} \hspace{0.05\linewidth}
    \subfloat[Annotated Mask\label{subfig:ExampleMask1}]{\includegraphics[width=0.45\linewidth]{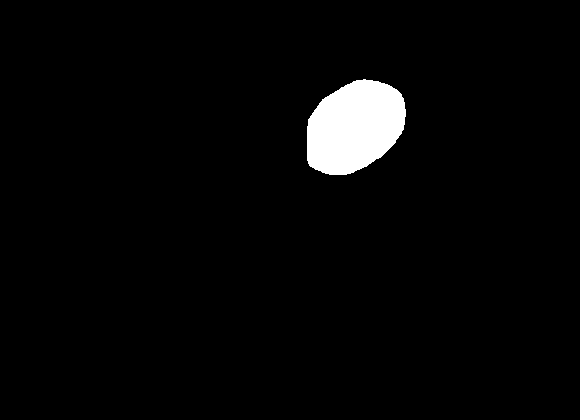}} \\
    \subfloat[Similar Ultrasound Image\label{subfig:ExampleImage2}]{\includegraphics[width=0.45\linewidth]{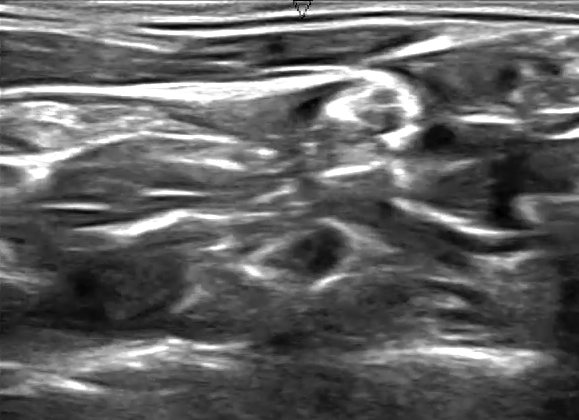}}\hspace{0.05\linewidth}
    \subfloat[Incoherent Mask\label{subfig:ExampleMask2}]{\includegraphics[width=0.45\linewidth]{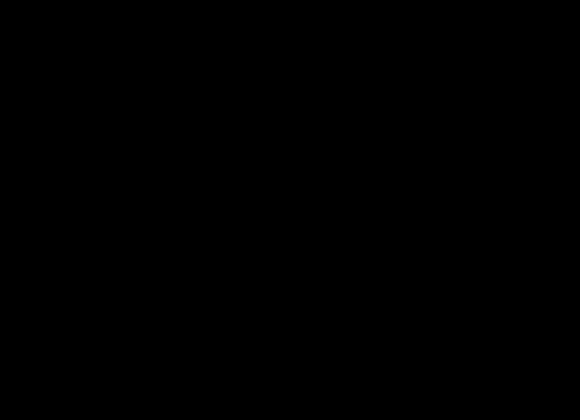}}
    \caption{(a,b) An example of an ultrasound image from the neck containing the brachial plexus nerve region with the corresponding annotated ground truth highlighting this region. (c,d) An example of similar ultrasound images with a complete different annotated brachial plexus nerve region.}
    \label{fig:example_images} 
\end{figure}

\subsection{Data}\label{section:data}
We use the data presented in \cite{KaggleDataset}. This dataset consisted of US images of the neck with their corresponding masks, as can be seen in Figure \ref{fig:example_images}. The images originate from $47$ different subjects, with 119-120 images per subject. All images had a resolution of 420x580 pixels. Trained volunteers annotated the binary masks indicating the BP. 

The dataset was found to contain contradictory annotations of close matching images. An example of this is shown in Figure \ref{fig:example_images}. A second dataset, here we call it filtered data, was constructed, in which the close matching samples without an annotated BP were removed.
Both the filtered and non-filtered data will be tested in the experiments. The dataset is unbalanced, with the no-BP class over-represented. The number of images in each class for both datasets are shown in Table \ref{table:Nr_images}. 

The data is then downsampled (128x128 in the CNN and 96x96 in the segmentation models). The images were mean-centered with a standard deviation of 1.  
In all experiments, 5-fold cross validation was used for the model evaluation. For each fold, the train data was randomly divided into a train and validation set by a 80\%-20\% split. To increase the number of samples used for training the CNN, we augmented training data by 2500 extra samples using affine transformations (scaling, shearing, rotating and reflecting) on randomly selected samples.

\renewcommand{\arraystretch}{1.2}
\begin{table*}[bth]
\caption{Experiment results (DSC) on filtered and non-filtered data.}
\label{table:DSC_Results}
\centering
\begin{tabular}{m{3cm} | m{1.5cm}  m{1.5cm} | m{1.5cm}  m{1.5cm} }
\hline
& \multicolumn{2}{c|}{\textbf{Non-filtered data}} & \multicolumn{2}{c}{\textbf{Filtered data}} \\
& \textit{U-net} & \textit{M-net} &  \textit{U-net} & \textit{M-net}\\
\hline
{No classification}      & 0.53 \textpm\ 0.06 & 0.59 \textpm\ 0.07 & 0.60 \textpm\ 0.08 & 0.65 \textpm\ 0.08 \\
{Hybrid model}           & 0.72 \textpm\ 0.01 & 0.67 \textpm\ 0.05 & 0.83 \textpm\ 0.02 & 0.79 \textpm\ 0.05 \\
{Perfect classification} & 0.92 \textpm\ 0.00 & 0.83 \textpm\ 0.10 & 0.93 \textpm\ 0.00 & 0.86 \textpm\ 0.09 \\
\hline
\end{tabular}
\end{table*}

\subsection{Experimental setup}

To investigate the effect of employing the hybrid model on the performance of segmenting BP region, three experiments were conducted:
\subsubsection*{Experiment 1: No classification} 
In the first experiment, the segmentation models were trained on the data without any prior classification model. In this experiment, training data contains US images with and without BP regions.
\subsubsection*{Experiment 2: hybrid model}
In the second experiment, the CNN architecture was used to identify images with BP. Then, the segmentation models (U-net and M-net) were trained on positive images only, and tested on the images that were classified by the CNN as BP.

\subsubsection*{Experiment 3: perfect classification model}
In the final experiment, a perfect classification model was mimicked by manually discarding all images without an annotated mask from data, and the segmentation models were trained and tested to these sets. This experiment was done to be able to discuss the potential of a hybrid model with a better performing classification model.

\subsection{Evaluation}
For evaluation of the classification results, two metrics were used in this research: the F1-score and the accuracy metric (see Equations \ref{eq:Acc} and \ref{eq:F1}). The accuracy metric is more suited for balanced datasets, whereas the F1-score is more suited for imbalanced datasets \cite{Tharwat2020}. Since the dataset was somewhat imbalanced, as discussed in Section \ref{section:data}, both metrics were evaluated.

For evaluation of the segmentation results, the dice similarity coefficient (DSC)  was used (see Equation \ref{eq:DSC}). DSC makes a pixel-wise comparison between the  ground truth and the corresponding prediction. The DSC ranges between 0 and 1, where 0 indicates there is no overlap, and 1 represents a perfect overlap.

All acquired data was tested for normality with the Shapiro-Wilk test~\cite{shapiro1965analysis}. The models were then tested for different means with a two sided T-test.

\begin{equation}\label{eq:Acc}
    {Accuracy} = \frac{{TP} + {TN}}{{TP} + {FP} +{TN} + {FN}}
\end{equation}
\begin{equation}\label{eq:F1}
    F1 = \frac{2 {TP}}{2{TP}+{FP}+{FN}}
\end{equation}
\begin{equation}\label{eq:DSC}
    DSC = \frac{2 \cdot |M\cap GT|}{|M| + |GT|}
\end{equation}

%% file: chapters/results.tex
\begin{figure}[tb]
    \centering
    \includegraphics[width=0.95\linewidth]{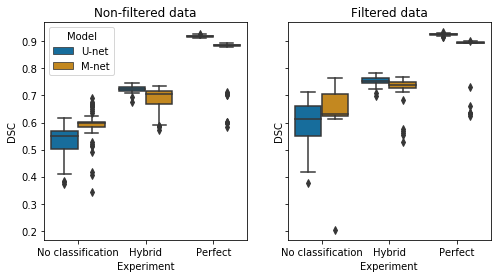}
    \caption{Results of the different experiments visualized in boxplots.}
    \label{fig:Boxplots}
\end{figure}

The CNN achieved an average F1-score of 0.72 \textpm\ 0.01 and an accuracy score of 0.77 \textpm\ 0.01 on the non-filtered data, whereas on the filtered data, an average F1-score of 0.82 \textpm\ 0.02 and an accuracy score of 0.87 \textpm\ 0.01 were found.

The DSCs of the experiments are listed in Table \ref{table:DSC_Results}. In the `no classification' experiment, M-net outperforms U-net in both the non-filtered and the filtered data ($p < .001, p = .002$ respectively). M-net and U-net both perform better on the filtered data, compared to the non-filtered data (both $p < .001$). 

From experiment 2, it was found that both hybrid architectures performed better in both datasets compared to the first experiment, as can be seen in Figure \ref{fig:Boxplots}. Both models still perform better on the filtered data, compared to the non-filtered data. The U-net variant has shown the most improvement, and now performs better than M-net in both datasets (all $p < .001$). In U-net, the variance has dropped as results became more stable. 

The third experiment shows an even better performance for both models in both datasets. Again, U-net has improved the most. The variance of U-net has decreased even more. This is not the case for M-net, where the variance has risen. Figure \ref{fig:Boxplots} shows the outliers (determined with the 1.5xIQR rule) in the M-net performance, explaining the large variance. Both models still perform better on the filtered data, however the discrepancy in performance between the two data sets is less compared to the first experiment.

%% file: chapters/discussion.tex
Our experimental results revealed the potential of using hybrid models. The `hybrid model' experiment showed an increase in performance of all models when prior classification is added, even when the classification is not perfect. The `perfect classification' experiment showed that promising results can be achieved when a perfect classification method can be approached or reached. It also showed that results will keep increasing as the classification becomes more accurate. 
 
U-net seems to have benefited more from the hybrid solution than M-net. This indicates that the M-net variant can better handle the large number of images without an annotated BP region, which can be attributed to the more densely supervised nature of the architecture \cite{Abraham2019}.

The difference between the performance on the non-filtered data and the filtered data increased for the hybrid variant of both U-net and M-net. The incoherent data may have caused errors in the learning process of the models. In the hybrid model, this would happen in both the classification and segmentation models, which would have caused the error to propagate. This propagating error could have been the cause of the increased difference in performance between filtered and non-filtered data.
This difference decreased in the third experiment. Since a perfect classification model was simulated, no images with a fully negative ground truth were sent to the segmentation models. This means that no incoherent data was being fed to the segmentation models when using the non-filtered dataset, since only incoherent negatives were removed in the filtered dataset. This can explain the reduction in performance difference between filtered and non-filtered data in the second experiment.

For the U-net model, the variance decreased when using the hybrid model, indicating a more stable model. Since most images without an annotated mask are filtered out by the CNN, more homogeneous data is fed to the segmentation model. This increases the likelihood for the segmentation model to recognize the BP regions. 

The results of M-net in our first experiment do not correspond with the results presented in \cite{Abraham2019}, with an average DSC of 0.59 in this study, compared to the reported 0.88 in \cite{Abraham2019}. Discrepancies between results can be expected since there are differences in methods. For one, the images used for testing are unknown, since each study created an own test set from the original data. Furthermore, the authors did not provide information about their test methods, making it harder to compare results. 

In future research, improving the classification model performance should be considered as an important step, since the perfect classification experiment showed the potential for much better segmentation performance. Moreover, merging the models into a single `chain' could improve time performance.

%% file: chapters/conclusion.tex
The goal of this research was to improve image segmentation of the BP nerve structure from US images, which was done by employing a hybrid deep neural network model. A CNN was used as the classification model, and U-net and M-net were employed as segmentation models. Three experiments were conducted on two datasets with filtered and non-filtered data. The first experiment did not use any prior classification, the second experiment used the CNN for classification, and in the third experiment, a perfect classification model was simulated by removing all negative samples from data. From these experiments, it could be concluded that performance significantly improved when using a hybrid model, and even more when using a `perfect classification' model, thus showing the promising possibilities of using hybrid model for accurate BP segmentation.